\documentclass[12pt]{article}
\usepackage{graphicx}
\usepackage{epsfig}
\usepackage{amsfonts}
\usepackage{amssymb}
\usepackage{amsmath}
\usepackage{mathrsfs}

\newcommand{\rom}[1]{\mathrm{#1}}

\newcommand{\beq}{\begin{equation}}
\newcommand{\eeq}{\end{equation}}
\newcommand{\be}{\begin{equation}}
\newcommand{\ee}{\end{equation}}
\newcommand{\beqa}{\begin{eqnarray}}
\newcommand{\eeqa}{\end{eqnarray}}
\newcommand{\beqar}{\begin{eqnarray*}}
\newcommand{\eeqar}{\end{eqnarray*}}

\newcommand{\reef}[1]{(\ref{#1})}

\numberwithin{equation}{section}

\begin{document}

\setlength{\unitlength}{1mm}

\thispagestyle{empty}
\begin{center}
{\bf \Large A black ring with a rotating 2-sphere}\\

\vspace*{1.4cm}

{\bf Pau Figueras}

\vspace*{0.3cm}

{\it Departament de F{\'\i}sica Fonamental, and}\\
{\it C.E.R. en Astrof\'{\i}sica, F\'{\i}sica de Part\'{\i}cules i Cosmologia,}\\
{\it Universitat de Barcelona, Diagonal 647, E-08028 Barcelona, Spain}\\[.3em]

\vspace*{0.2cm}
{\tt pfigueras@ffn.ub.es}

\vspace{.8cm} {\bf ABSTRACT}
\end{center}
We present a solution of the vacuum Einstein's equations in five dimensions corresponding to a black ring with horizon topology $S^1\times S^2$ and rotation in the azimuthal direction of the $S^2$. This solution has a regular  horizon up to a conical singularity, which can be placed either inside the ring or at infinity. This singularity arises due to the fact that this black ring is not balanced. In the infinite radius limit we correctly reproduce the Kerr black string, and taking another limit we recover the Myers-Perry black hole with a single angular momentum.

\noindent

\vfill \setcounter{page}{0} \setcounter{footnote}{0}
\newpage

\tableofcontents

\section{Introduction}
The discovery of the black ring by \cite{ER} showed that the four-dimensional black hole theorems \cite{classic} do not have a simple extension to higher dimensions.  Black rings have a horizon with topology $S^1\times S^2$ while the Myers-Perry (MP) black hole \cite{Myers}  in five dimensions has a horizon with the topology of an $S^3$. On the other hand, in four dimensions, the black hole theorems state that the only allowed horizon topology is that of an $S^2$. Moreover, it was shown in \cite{ER} that both the black hole and the black ring  can carry the same conserved charges, namely the mass and a single angular momentum, and hence there is no uniqueness theorem in five dimensions. Further studies on black rings showed that they can also carry nonconserved charges \cite{RE}, which is a completely novel feature with respect to spherical black holes. The fact that black rings can have \textit{dipole} charges gives rise to infinite continuous nonuniqueness at the classical level since the parameters describing these charges can be varied continuously without altering the conserved charges.

If string theory is the correct theory of quantum gravity, it should account for the microscopic states of both spherical black holes and black rings. Therefore, black rings constitute an interesting test ground for string theory. A step towards the understanding of black rings in string theory was taken in \cite{EE}, following \cite{Elvang}, where it was found that  black rings were related to supertubes \cite{MT,EMT}, which are well-known objects within the string theory framework. The recent discovery of supersymmetric black rings \cite{EEMR1} (see also \cite{EEMR2,BW,GG}) and the generalization to non-BPS black rings with three charges and three dipoles \cite{EEF}, has provided further evidence of this relationship. Both the BPS and the non-BPS black rings with three charges and three dipoles have two angular momenta $J_\psi$  and $J_\phi$.  In fact, \cite{ER} already conjectured the existence a neutral black ring with two independent angular momenta. Moreover,  in \cite{EEF} a further  conjecture was made about the existence of a family of non-supersymmetric black rings with nine parameters $(M,J_\psi,J_\phi,Q_{1,2,3},q_{1,2,3})$.  Recently \cite{Larsen} conjectured the existence of an even larger family of black rings, which would depend on twenty-one parameters, up to duality transformations. Motivated by these conjectures, in this paper we present a neutral black ring with rotation in the azimuthal direction of the $S^2$, which from now on we call $\phi$. Recall that the black ring of  \cite{ER} has rotation on the plane of the ring along the direction of the $S^1$, which we call $\psi$. Our solution should be the $J_\psi \to 0$ limit of the more general yet to be found doubly spinning black ring. The solution we present in this paper  provides new evidence in favour of the existence of this neutral  black ring with two independent angular momenta. 

Ref.  \cite{Mishima}  has independently constructed a  black ring with rotation in the $S^2$.  It is claimed  that their solution reproduces, in appropriate limits, the Kerr black string and the MP black hole with a single angular momentum, as  does ours and hence both solutions may be equivalent. However, this is not evident since the coordinates used in \cite{Mishima} are rather involved. Instead, we present the solution in essentially the same coordinates as in \cite{RE} so that the  connections with the previously known black ring solutions are immediately apparent. Using these coordinates we study all its properties, including the different limits that this solution  meets. Specifically, we find that our solution has a regular horizon except for the conical excess singularity inside the ring which prevents it from collapsing. We could also choose to place the conical defect at infinity, which would equally stabilize the configuration, but the resulting spacetime would not be asymptotically flat. The presence of this singularity can be understood intuitively;  a black ring can be constructed by taking a black string, identifying its ends so as to form an $S^1$ and adding angular momentum in the direction of the $S^1$  in order to compensate the self-attraction and the tension of the string. In our case, since the angular momentum is on the plane orthogonal to  the ring, it does not balance the configuration and a conical singularity inside the ring is required. Our solution has three independent parameters, namely $R$, which is roughly the radius of the ring, $\lambda$, which plays a similar role  as the $\nu$ parameter in the static ring \cite{Weyl}\footnote{Note that the C-metric coordinates used in \cite{Weyl,ER} are different from the ones used in \cite{EE,Hong}, which in turn are different from those in \cite{RE}. Throughout this paper we use the same coordinates as ref. \cite{RE}.} and is related to the conical singularity, and a new parameter $a$ with dimensions of length, which is associated to the angular momentum. Taking the limit $a\to 0$ of our solution, one obtains the static ring \cite{Weyl} in the form given in \cite{RE}.

The rest of the paper is organized as follows. In section \ref{sec:solution} we present the solution and compute the main physical properties. Also, we study the  horizon geometry and derive the Smarr relation. In section \ref{sec:limits} we study the different limits of our solution. Specifically, taking infinite radius limit we obtain the Kerr black string. Moreover, we also study the limit which connects our solution with the MP black hole with a single angular momentum. The conclusions are presented in section \ref{sec:conclusions}.

\section{The black ring with a rotating $S^2$}
\label{sec:solution}
In this section we present the black ring with rotation on the $S^2$. The solution has been obtained by educated guesswork, from limits it would be expected to reproduce. It looks rather similar to the other neutral black ring solutions, which involve simple polynomic functions of $x$ and $y$.  Here we study the physical properties of the solution and the  horizon geometry. 

%The metric is
%\beqa
%  \nonumber
%  ds^2 &=& 
%  -\frac{1+\lambda y + (a\, x\, y/R)^2}{1+\lambda x + (a\, x\, y/R)^2} 
%  \left[ dt - 
%   \frac{\lambda\, a\, y \, (1-x^2)}
%        {1+\lambda y + (a\, x\, y/R)^2} \, d\phi\right]^2 \\[2mm]
% && ~~
%  + \frac{R^2}{(x-y)^2} 
%  \Bigg[ 
%    - \frac{1+\lambda x + (a\, x\, y/R)^2}
%           {1+\lambda y + (a\, y/R)^2} \frac{dy^2}{1-y^2}
%    - (1-y^2) \Big[ 1+\lambda x + (a\, x/R)^2\Big] \, d\psi^2 
%  \\[2mm] \nonumber
%  && ~~
%   + \frac{1+\lambda x + (a\, x\, y/R)^2}{1+\lambda x + (a\, x/R)^2}
%     \frac{dx^2}{1-x^2}
%   + (1-x^2) 
%     \frac{\Big[1+\lambda y + (a\, y/R)^2\Big]
%           \Big[1+\lambda x + (a\, x\, y/R)^2\Big]}
%          {1+\lambda y + (a\, x\, y/R)^2} \, d\phi^2\Bigg] \, .
%\eeqa
\subsection{The solution}

The metric for the $\phi$-spinning ring is
\beqa
  \nonumber
  ds^2 &=& -\frac{H(\lambda,y,x)}{H(\lambda,x,y)} 
  \left[ dt - 
   \frac{\lambda\, a\, y \, (1-x^2)}
        {H(\lambda,y,x)} \, d\phi\right]^2  \\[2mm]
 && ~~
  + \frac{R^2}{(x-y)^2}~H(\lambda,x,y) 
  \Bigg[ 
    - \frac{dy^2}{(1-y^2)F(\lambda,y)}
    - \frac{(1-y^2) F(\lambda,x)}{H(\lambda,x,y)}~d\psi^2 \label{eqn:phispin} 
  \\[2mm] \nonumber
  && ~~\hspace{4.2cm}
   + \frac{dx^2}{(1-x^2)F(\lambda,x)}
   + \frac{(1-x^2)F(\lambda,y)}{H(\lambda,y,x)}~d\phi^2
   \Bigg]\, ,
   \nonumber
\eeqa
where
\beqa
  F(\lambda,\xi) = 1 + \lambda\, \xi + \left(\frac{a\, \xi}{R}\right)^2
  \;,\hspace{1cm}
  H(\lambda,\xi_1,\xi_2) = 1 + \lambda\, \xi_1 + 
    \left(\frac{a\, \xi_1\, \xi_2}{R}\right)^2 \, .
\eeqa
The coordinates are the same as those used in \cite{RE}, and their ranges are
\beq
  1 < x < -1 \, , ~~~~~
  -\infty <y < -1 \, .
\eeq
Recall that in this set of coordinates, $\psi$ parametrizes an $S^1$ and $(x,\phi)$ an $S^2$, with $x\sim \cos\theta$, where $\theta$ is the polar angle, and $\phi$  the azimuthal angle. From \eqref{eqn:phispin} one sees that the black ring is  rotating along the azimuthal direction of the $S^2$, while the previously known neutral black ring \cite{ER} rotates along the $\psi$ direction of the $S^1$.
 
 Throughout this paper we assume that
\beq
  \frac{2 a}{R} < \lambda < 1 + \frac{a^2}{R^2} \label{eqn:bound}
\eeq
where the upper bound garantees that both $g_{xx}$ and $g_{\phi\phi}$ are positive and hence there are no closed time-like curves \cite{EEMR1,EEMR2}. As we will show later on, the lower bound is required for there to be a horizon.  Setting $a=0$ in \eqref{eqn:phispin} gives the static ring first found in \cite{Weyl}.

As in other black ring solutions, $R$ corresponds roughly to the radius of the black ring, whereas $a$ is a new parameter with dimensions of length which is related to the angular momentum, as will be seen from the expressions of the conserved charges. In fact, in section \ref{sec:limits} it will become  manifest that  $a$ is roughly equivalent to the rotation parameter of the four dimensional Kerr black hole.

In order to avoid a conical singularity at  $x=-1$ and $y=-1$, we have to identify the periods of the angular coordinates $\psi$ and $\phi$ according to
\beq 
  \Delta \phi = \Delta \psi = \frac{2\pi}{\sqrt{1 - \lambda + a^2/R^2}} \, . \label{eqn:delta1}
\eeq
Also, in order for the orbits of $\partial_\phi$ to close off smoothly at $x=1$ we must require  
\beq
  \Delta \phi = \frac{2\pi}{\sqrt{1 + \lambda + a^2/R^2}}\;,
\eeq
which cannot be satisfied at the same time as \reef{eqn:delta1} for
nonzero $\lambda$. For the asymptotically flat ring, eq.~\reef{eqn:delta1}
holds and there is a conically singular disk at $x=1$ supporting the
ring. From now on we will stick to this case.

 Asymptotic infinity is at $x,y \to -1$. Now, defining canonical angular variables,
\beqa
\tilde\psi=\frac{2\pi}{\Delta\psi}~\psi\;,&& \tilde\phi=\frac{2\pi}{\Delta\phi}~\phi\;,
\eeqa
with $\Delta\psi$ and $\Delta\phi$ given in \reef{eqn:delta1}, and performing the coordinate transformation
\beqa
\zeta=R~\frac{\sqrt{2(-1-y)}}{x-y}\;,&& \eta=R~\frac{\sqrt{2(1+x)}}{x-y}\;,
\eeqa
the metric \reef{eqn:phispin} takes the manifestly flat form
\beq
ds^2\sim-dt^2+\zeta^2 d\tilde\psi+d\zeta^2+\eta^2d \tilde\phi^2+d\eta^2\;.
\eeq
The ADM mass and angular momentum are given by
\beqa
M&=&\frac{3\pi R^2}{4G}\frac{\lambda}{1-\lambda+ a^2/R^2}\;,\nonumber\\
J_\phi&=&-\frac{\pi R^2}{G}\frac{\lambda a}{\left(1-\lambda+ a^2/R^2\right)^{3/2}}\;.
\eeqa
Again, for $a=0$ the angular momentum vanishes and the mass coincides with that of the static ring given in \cite{RE}.

The Killing vector field $\partial_t$ becomes spacelike at the least negative zero of
$H(\lambda,y,x)$, $y_e$; this point defines the ergosurface. Similarly, the zeros of $F(\lambda,y)$ 
\beq
  y_i = \frac{-\lambda-\sqrt{\lambda^2 - 4 a^2/R^2}}{2 a^2/R^2}
  \, , ~~~~~~
  y_h = \frac{-\lambda+\sqrt{\lambda^2 - 4 a^2/R^2}}{2 a^2/R^2}\;,
  \label{eqn:horizon}
\eeq
determine the locations of the inner and outer horizons, respectively. In the limit $a \to 0$, we have $y_h \to -1/\lambda$, which is the location of the horizon for the static black ring, and $y_i \to -\infty$, for which the static ring has a curvature singularity. Both the ergosurface and the horizon have topology $S^1\times S^2$, but we will come back to this point in the next subsection. Moreover, it can be checked that the Killing vector field $\partial_\phi$ remains spacelike throughout the ergoregion, which is the region $y_h<y<y_e$. As in the four-dimensional Kerr black hole, in the $\phi$-spinning ring the ergoregion coincides with the horizon at the poles of the $S^2$ $(x=\pm 1)$. Finally, from \reef{eqn:horizon} it is clear that we must impose $\lambda > 2a/R$ so as to have a  horizon.

\subsection{ Horizon geometry}
\label{s:phihorizon}

The Kretschmann invariant is finite at $y=y_i, y_h$, but blows up when 
$H(\lambda,x,y)=0$. To find coordinates that are good on the horizon we proceed as in \cite{EEMR1,EEMR2} and define $\bar y = y - y_h$. Consider the following change of coordinates
\beq
  dt = dv + \frac{A}{\bar y} \, d\bar y \, , ~~~~
  d\phi = d\chi + \frac{B}{\bar y} \, d\bar y \, ,
\eeq
where $A$ and $B$ are determined by requiring that the solution in the coordinates $(v,x,\bar y,\psi,\chi)$ is analytic at $\bar y=0$. When we expand near $\bar y=0$, we find that $g_{v\bar y}$ and $g_{\chi \bar y}$ diverge as $1/\bar y$. Both divergences can be removed by imposing $A = -B \, \lambda \, R^2/(a \, y_h)$. This choice also removes the $1/\bar y^2$-divergence in $g_{\bar y\bar y}$. Then  $g_{\bar y\bar y}$ diverges as $1/\bar y$ unless we impose
\beq
  B^2 = \frac{y_h^2~a^2/R^2}{(y_h^2-1)(\lambda^2-4 a^2/R^2)}\; .
\eeq 
The metric is now analytic at $\bar y=0$, and its form on a spatial cross-section of the horizon is  rather simple:
\beq
  ds^2_\rom{H} = \frac{R^2}{(x-y_h)^2} 
    \left[   
    (y_h^2-1) F(\lambda,x) \, d\psi^2
   + \frac{H(\lambda,x,y_h)}{(1-x^2)F(\lambda,x)}~dx^2 \right]
   + (1-x^2) \frac{\lambda^2 R^2}{H(\lambda,x,y_h)} \, d\chi^2 \;.
   \label{eqn:area}
\eeq
This expression shows that the horizon is regular, up to the conical singularity, and that its topology is indeed $S^1\times S^2$, where the $S^1$ is parametrized by the angle $\psi$ and the $S^2$ by $(x,\phi)$. Using \reef{eqn:area} we can compute the horizon area
\beq
  {\cal A}_\rom{H} = \frac{8 \pi^2 R^3 \lambda}
  {(1-\lambda +a^2/R^2)\sqrt{y_h^2-1}} \, .
\eeq
For $a=0$ this agrees with the $\nu = \lambda$ limit of the $\psi$-rotating black ring.

The Killing vector field
\beq
\partial_v -\frac{a\, y_h}{\lambda R^2}~\partial_\chi
\eeq
becomes null at $y=y_h$, and hence we can identify it as the null generator of the horizon. The angular velocity of the horizon and the temperature are given by
\beqa
\Omega_\phi&=&-\frac{\sqrt{1-\lambda+a^2/R^2}\big[\lambda-\sqrt{\lambda^2-4a^2/R^2}\big]}{2\lambda a}\;,\\
T_{\rom{BH}}&=&\frac{\sqrt{(\lambda^2-4a^2/R^2)(y_h^2-1)}}{4\pi R \lambda}\;.
\eeqa
Using these results, a straightforward calculation shows that the $\phi$-spinning ring also satisfies a Smarr relation,
\beq
M=\frac{3}{2}\left[\frac{1}{4G}{\cal A}_\rom{H}T_{\rom{BH}}+\Omega_\phi J_\phi\right]\;.
\eeq
This relationship holds in the case in which the solution is asymptotically flat, that is, the conical defect is inside the ring.

\section{Limits}
\label{sec:limits}

\subsection{Infinite radius limit}
In order to obtain the infinite radius limit we have to send $R \to \infty$ while keeping fixed 
\beq
  r = - \frac{R}{y} \, , ~~~~
  \cos\theta = x \, , ~~~~
  2M = \lambda R \, , ~~~~
  \eta = R\psi \, .
\eeq
We then have 
\beq
  F(\lambda,x) \to 1 \, , ~~~~
  F(\lambda,y) \to \Delta/r^2 \, , ~~~~
  H(\lambda,x,y) \to \Sigma/r^2 \, , ~~~~
  H(\lambda,y,x) \to (\Sigma-2Mr) /r^2 \, , ~~~~
\eeq
with $\Delta =  r^2 - 2Mr +a^2$ and $\Sigma = r^2 + a^2 \cos^2\theta$.
In this limit the metric \reef{eqn:phispin} becomes the Kerr black
string,
\beqa
ds^2=-\left(1-\frac{2Mr}{\Sigma}\right)\left[dt+\frac{2Mr}{\Sigma-2Mr}a\sin^2\theta~d\phi\right]^2
	+\Sigma\left(\frac{dr^2}{\Delta}+d\theta^2\right)
	+\frac{\Delta \sin^2\theta}{1-\frac{2Mr}{\Sigma}}d\phi^2+d\eta^2\;.
\eeqa

In this limit, the singularity encountered at $H(\lambda,x,y)=0$ becomes the ring-like singularity of the Kerr spacetime.

\subsection{Myers-Perry black hole limit}
In order to recover the five-dimensional MP black hole with rotation in the $\phi$ direction from \reef{eqn:phispin}, we have to take a limit similar to the one described in \cite{RE}. We define new parameters 
\beq
m=\frac{2(R^2+a^2)}{1+a^2/R^2-\lambda}\,,\quad \alpha^2=\frac{4
a^2}{1+a^2/R^2-\lambda}\,,
\eeq
such that they remain finite as $\lambda\to 1+a^2/R^2$, $R\to 0$ and $a\to 0$. Furthermore, we perform a change of coordinates $(x,y)\to(r,\theta)$
\beqa
x&=&-1+\frac{2R^2\cos^2\theta}{r^2-(m-\alpha^2)\cos^2\theta}\,,\\
y&=&-1-\frac{2R^2\sin^2\theta}{r^2-(m-\alpha^2)\cos^2\theta}\,,
\eeqa
and rescale $\psi$ and $\phi$ so that in this limit they have the canonical periodicity $2\pi$,
\beq
(\tilde\phi,\tilde\psi)=\sqrt{\frac{2(R^2+a^2)}{m}}\;(\phi,\psi)\,.
\eeq
Then, we obtain the MP black hole with rotation in $\phi$
\beq
ds^2=-\left(1-\frac{m}{\Sigma}\right)\left(dt-
\frac{m\alpha\cos^2\theta}{\Sigma-m}d\tilde{\phi}\right)^2
+\Sigma\left(\frac{dr^2}{\Delta}+d\theta^2\right)
+\frac{\Delta\cos^2\theta}{1-\frac{m}{\Sigma}}d\tilde{\phi}^2 +r^2\sin^2\theta\, d\tilde{\psi}^2\;,
\eeq
where
\beq
\Delta=r^2-m+\alpha^2\,,\qquad \Sigma=r^2+\alpha^2\sin^2\theta\;.
\eeq

\section{Conclusions}
\label{sec:conclusions}
In this paper we have presented a black ring solution with rotation on the plane orthogonal to the ring. Although our solution has a horizon with topology $S^1\times S^2$, there is a conical singularity inside the ring that prevents it from collapsing.  This solution should be an intermediate step towards a more general black ring with two independent angular momenta.

Using \eqref{eqn:phispin} as a seed solution and following the procedure outlined in \cite{EE,EEF}, one can construct the  black ring with a rotating $S^2$ and carrying F1-P charges. In this charged black ring, which we do not present here, the conical singularity persists except in the extremal supersymmetric limit.  In this limit, in which the angular momentum disappears,  we find a configuration which can be interpreted as a continuous distribution of straight F1-P strings on a circle. We do not expect to be able to add more than two charges to the solution \eqref{eqn:phispin} without generating Dirac-Misner strings for the reasons discussed in \cite{EE}. Furthermore, it would  be very  interesting to construct the $\phi$-spinning ring carrying dipole charges. Even then, we do not expect to be  able to remove  the conical singularity since the dipole charges typically increase the tension of the string \cite{RE}. Adding three charges to this more general solution should result in a charged $\phi$-ring with dipole charges. We hope our results are of help towards a classification of the possible black holes of five-dimensional vacuum gravity. This is also a step in the construction of the seed solution that would generate the 9-parameter non-supersymmetric black rings conjectured in \cite{EEF} (and, further, the larger family conjectured in \cite{Larsen}), that should describe thermal excitations of three-charge supertubes. We hope to tackle these problems in the future.

\section*{Acknowledgments}

\noindent We thank Henriette Elvang and Roberto Emparan for their invaluable help in this work. We also thank Harvey Reall for useful comments and discussions. This work  was supported  by a FI scholarship from Generalitat de Catalunya, CICyT FPA2004-04582-C02-02  and European Comission RTN program under contract MRTN-CT-2004-005104.

\vfill

\newpage

%%%%%%%%%%%%%%%%%%%%%%%%%%%%%%%%%%%%%%%%%%%%%%%%%%%%%%%

\end{document}